\documentclass[conference]{IEEEtran}
\usepackage{cite}

\usepackage[dvipsnames,svgnames]{xcolor}
\usepackage{tikz,pgfplots}
\usetikzlibrary{arrows,decorations.markings}
\usepackage{graphicx}
\usepackage{psfrag}
\DeclareGraphicsExtensions{.xps}

\usepackage{epstopdf} 
\usepackage[cmex10]{amsmath}
\usepackage{amssymb}
\usepackage{amsthm}

\def\tr{\mathrm{tr}}

\newtheorem{lemma}{Lemma}

\newtheorem{remark}{Remark}
\newtheorem{assumption}{Assumption}

\usepackage[textwidth=30mm]{todonotes}

\newcommand{\vect}[1]{\mathbf{#1}}

\def\tr{\mathrm{tr}}

\def\vectorize{\mathrm{vec}}
\def\Htran{\mbox{\tiny $\mathrm{H}$}}
\def\Ttran{\mbox{\tiny $\mathrm{T}$}}
\def\CN{\mathcal{N}_{\mathbb{C}}} 
\def\taupu{\tau_{p}} 
\def\bphiu{\boldsymbol{\phi}} 

\begin{document}

\IEEEoverridecommandlockouts

\title{Fundamental Asymptotic Behavior of (Two-User) Distributed Massive MIMO\vspace{-.4cm}}
\author{
\IEEEauthorblockN{Luca Sanguinetti\IEEEauthorrefmark{1}, Emil Bj{\"o}rnson\IEEEauthorrefmark{2}, Jakob Hoydis\IEEEauthorrefmark{3} \bigskip
\vspace{-.4cm}}
\IEEEauthorblockA{\IEEEauthorrefmark{1}\small{Dipartimento di Ingegneria dell'Informazione, University of Pisa, Pisa, Italy}}
\IEEEauthorblockA{\IEEEauthorrefmark{2}\small{Department of Electrical Engineering (ISY), Link\"{o}ping University, Link\"{o}ping, Sweden}}
\IEEEauthorblockA{\IEEEauthorrefmark{3}\small{Nokia Bell Labs, Nozay, France.}}
\vspace{-0.8cm}}

\maketitle
\begin{abstract}
This paper considers the uplink of a distributed Massive MIMO network where $N$ base stations (BSs), each equipped with $M$ antennas, receive data from $K=2$ users. We study the asymptotic spectral efficiency (as $M\to \infty$) with spatial correlated channels, pilot contamination, and different degrees of channel state information (CSI) and statistical knowledge at the BSs. 
By considering a two-user setup, we can simply derive fundamental asymptotic behaviors and provide novel insights into the structure of the optimal combining schemes. In line with~\cite{BHS18A}, when global CSI is available at all BSs, the optimal minimum-mean squared error combining has an unbounded capacity as $M\to \infty$, if the global channel covariance matrices of the users are asymptotically linearly independent. This result is instrumental to derive a suboptimal combining scheme that provides unbounded capacity as $M\to \infty$ using only local CSI and global channel statistics. The latter scheme is shown to outperform a generalized matched filter scheme, which also achieves asymptotic unbounded capacity by using only local CSI and global channel statistics, but is derived following \cite{Neumann2018} on the basis of a more conservative capacity bound. 
\end{abstract}

\section{Introduction}


Massive MIMO refers to a wireless network technology where the base stations (BSs) are equipped with a very large number $M$ of antennas to serve a multitude of user equipments (UEs) by spatial multiplexing \cite{Marzetta2010,Larsson2014}. Exciting developments have occurred in the recent year. In industry, the technology has been integrated into the 5G New Radio standard \cite{Parkvall2017a}. In academia, the long-standing pilot contamination issue, which was believed to impose fundamental limitations \cite{Marzetta2010}, has finally been resolved \cite{BHS18A}. More precisely, UEs that use the same pilot sequence for channel estimation cause interference that first seemed impossible to suppress. However, when using the optimal minimum mean squared error (MMSE) combining/precoding scheme \cite{BHS18A} or the generalized matched filters \cite{Neumann2018}, the capacity grows unboundedly as $M\to \infty$. 
The key to prove these results was to utilize the spatial correlation of practical channels \cite{Gao2015b}, which gives the BS sufficient prior information to distinguish between UEs even when their channel covariance matrices have full rank.

Distributed Massive MIMO refers, in this paper, to a group of Massive MIMO BSs that jointly process the signals to/from a joint set of UEs \cite{Truong2013}. Each BS only needs local channel state information (CSI), from the UEs to itself, in order to function. The asymptotic performance limits of these networks, as $M \to \infty$, are less explored. 
A pilot-contamination precoding scheme for spatially uncorrelated channels is proposed in \cite{ashikhmin2012pilot}, as a way to achieve an unbounded capacity as $M \to \infty$. Only locally obtained channel estimates are used at each BS but their signals are jointly processed by inverting a matrix containing statistical channel coefficients, whose invertibility was not analytically proved in \cite{ashikhmin2012pilot}.

In this paper, we consider the uplink of a two-UE distributed Massive MIMO system with spatially correlated channels. Each BS is equipped with $M$ antennas and receives data from UE 1 and pilot-contaminated interference from UE 2. This setup is sufficient to prove and demonstrate our main results. In particular, we show rigorously under which conditions the system can achieve an unbounded capacity in the presence of pilot contamination. We consider two cases with different amounts of information at the BSs: 1) global CSI and channel statistics; 2) local CSI and global statistics. In both cases, we generalize known schemes and derive new mathematical formulations that shed light on the different ways that an unbounded asymptotic capacity can be achieved in distributed Massive MIMO. For the second case, we also provide a novel distributed and low complexity receive combining scheme. 

\subsubsection*{Notation} The Frobenius norm of a matrix $\vect{X}$ is denoted by $\| \vect{X} \|_F$. 
We use $\CN({\bf 0},{\bf R})$ to denote the circularly symmetric complex Gaussian distribution with covariance matrix ${\bf R}$. The $N \times N$ identity matrix is denoted by $\vect{I}_N$. 
 We use $a_n \asymp b_n$ to denote $a_n -b_n \to_{n\to \infty}0$ almost surely (a.s.) for two sequences of random variables $a_n$, $b_n$. With a slight abuse of notation, ${\bf A}\asymp{\bf B}$ denotes that the matrices ${\bf A},{\bf B}\in\mathbb{C}^{M\times M}$ are asymptotically equivalent, in the sense that $\lim_{M \to \infty} \| {\bf A} - {\bf B} \|_F=0$. We use $\vectorize({\bf A})$ and $\vectorize^{-1}({\bf A})$ to denote the vectorization of a matrix ${\bf A}\in\mathbb{C}^{M\times N}$ and its inverse operation. We denote ${\bf e}_k$ the $k$th vector of the canonical basis.
 

\begin{figure*}[ht]
\setcounter{equation}{6}
\begin{align}
\gamma_{1} 
&= \frac{ \left|  \sum\limits_{n=1}^N{(\vect{v}_{1}^n)}^{\Htran} \hat{\vect{h}}_{1}^n \right|^2  }{ {\mathbb{E}}\bigg\{\left| \sum\limits_{n=1}^N{(\vect{v}_{1}^n)}^{\Htran} {\vect{h}}_{2}^n\right|^2  - \left|  \sum\limits_{n=1}^N{(\vect{v}_{1}^n)}^{\Htran} \tilde{\vect{h}}_{1}^n  \right|^2  + \frac{1}{\rho}\sum\limits_{n=1}^N|| \vect{v}_{1}^n||^2\bigg|\{\hat{\vect{h}}_{1}^n\}, \{\hat{\vect{h}}_{2}^n\} \bigg\}}=\frac{ \left|  \sum\limits_{n=1}^N{(\vect{v}_{1}^n)}^{\Htran} \hat{\vect{h}}_{1}^n \right|^2  }{ \left|  \sum\limits_{n=1}^N{(\vect{v}_{1}^n)}^{\Htran} \hat{\vect{h}}_{2}^n \right|^2 + \sum\limits_{n=1}^N {(\vect{v}_{1}^n)}^{\Htran} {\bf Z}^n\vect{v}_{1}^n}\label{eq:gamma1_UL}
\end{align}
\vspace{-0.1cm}
\hrule\vspace{-0.5cm}
\end{figure*}
\setcounter{equation}{0}
\section{System Model}
\label{section:two-user}

We consider a two-user uplink scenario, where $N$ distributed BSs, each equipped with $M$ antennas, receive data from UE~$1$ and pilot-contaminated interference from UE~$2$ (and vice-versa). Denote by 
${\bf h}_{k}^{n}  \in \mathbb{C}^{M}$ the channel from UE~$k$ to BS $n$. We consider a correlated Rayleigh block fading model $\vect{h}_{k}^{n} \sim \CN \left( \vect{0}, \vect{R}_{k}^{n}  \right)$ for $k=1,2$ where ${\vect{R}_{k}^{n} \in \mathbb{C}^{M\times M}}$, with $\tr ( \vect{R}_{k}^{n} ) > 0$, is the channel covariance matrix. The Gaussian distribution models the small-scale fading whereas the covariance matrix $\vect{R}_{k}^{n}$ describes the large-scale fading, including pathloss and spatial correlation. 
We assume that channel vectors of different BSs are independent, thus ${\mathbb{E}}\{\vect{h}_{k}^{n}{(\vect{h}_{k}^{i})}^{\Htran}\} = {\bf 0}$ for $i\ne n$. This is a reasonable assumption since the BSs are spatially distributed in the network. For later convenience, we define also the global covariance matrix $\vect{R}_{k} = {\rm{diag}}\left(\vect{R}_{k}^1,\ldots,\vect{R}_{k}^N\right) \in \mathbb{C}^{NM\times NM}$,
which is block-diagonal. We assume that the channel covariance matrices $\{\vect{R}_{k}^n;k=1,2\}$ are locally available at each BS $n$; see \cite{BSD16A,NeumannJU17,UpadhyaJU17,CaireC17a} (among others) for practical  methods for covariance matrix estimation.

We assume that the BSs and UEs are perfectly synchronized and operate according to a 
protocol with a data transmission phase and a pilot phase for channel estimation.

\subsection{Data Transmission Phase}

During the uplink data transmission, the received complex baseband signal ${\bf y}^n \in \mathbb{C}^{M}$ at BS $n$ is given by
\begin{align}
\vect{y}^n= \sqrt{\rho} \vect{h}_{1}^n x_1 + \sqrt{\rho} \vect{h}_{2}^n x_2 + \vect{n}^n
\end{align}
where $x_k\sim\CN(0,1)$ is the information-bearing signal transmitted by UE~$k$, $\vect{n}^n\sim \CN(\vect{0},{\bf I}_M)$
is the normalized independent receiver noise, and $\rho$ is the signal-to-noise ratio (SNR). BS $n$ processes the signal from UE~$k$ using a receive combining vector $\vect{v}_k^n\in \mathbb{C}^{M}$ to obtain the scalar ${(\vect{v}_k^n)}^{\Htran}\vect{y}^n$. The latter are then combined between the BSs to obtain $\sum_{n=1}^N{(\vect{v}_k^n)}^{\Htran}\vect{y}^n$, which is  used to decode $x_k$ for $k=1,2$.

\subsection{Channel Estimation Phase}

The combining vector $\vect{v}_1^n$ is computed on the basis of the CSI available at BS $n$, and acquired from pilot transmission. Both UEs use the same $\taupu$-length pilot sequence $\bphiu \in \mathbb{C}^{\taupu}$ with elements such that $\| \bphiu \|^2  = \bphiu^{\Htran} \bphiu  = \tau_p$, since we want to study the fundamental impact of pilot contamination. The received uplink signal $\vect{Y}^{n} \in \mathbb{C}^{M\times \taupu}$ at {BS} $n$ is
\begin{align}\label{eq:observation_vector}
\vect{Y}^{n}= \sqrt{\rho_{\rm{tr}}} \vect{h}_{1}^n \bphiu^{\Ttran} + \sqrt{\rho_{\rm{tr}}}  \vect{h}_{2}^n \bphiu^{\Ttran} + \vect{N}^{n}
\end{align}
where $\rho_{\rm{tr}} = \rho \tau_p$ is the pilot SNR and $\vect{N}^{n} \in \mathbb{C}^{N\times \taupu}$ is the normalized receiver noise with all elements independently distributed as $\CN(0,1)$. The matrix $\vect{Y}^{n}$ is the observation that {BS} $n$ utilizes to estimate the channels ${\bf h}_k^n$ for $k=1,2$. 
Since ${\bf h}_k^n\sim \CN \left( \vect{0}, \vect{R}_{k}^{n}  \right)$ is a realization of a random variable whose distribution is known, the {MMSE}  estimator is used.

\begin{lemma}\cite[Th. 3.1]{massivemimobook}\label{theorem:MMSE-estimate_h_jli}
The {MMSE} estimator of $\vect{h}_{k}^{n}$ for $k=1,2$, based on the observation $\vect{Y}^{n} $, is
\begin{equation} \label{eq:MMSEestimator_h}
\begin{split}
\!\!\hat{\vect{h}}_{k}^n  =  \frac{1}{ \sqrt{\rho_{\rm{tr}}} }\vect{R}_{k}^n
 {({\bf{Q}}_{\rm{tr}}^n)}^{-1}\vect{Y}^{n} \bphiu^{*}
\end{split}
\end{equation}
where 
\begin{equation}\label{eq:covariance_matrix_pilot_phase}
{\bf{Q}}_{\rm{tr}}^n = \vect{R}_{1}^n +  \vect{R}_{2}^n + \frac{1}{ \rho_{\rm{tr}}} \vect{I}_{M}.
\end{equation}
\end{lemma}
Since the two UEs use the same pilot, the estimates $\hat {\bf h}_k^n$ for $k=1,2$ are strongly correlated; the correlation matrix is ${\mathbb{E}}\{\hat{\vect{h}}_{2}^n(\hat{\vect{h}}_{1}^n)^{\Htran}\}  = \vect{R}_{2}^n
{({\bf{Q}}_{\rm{tr}}^n)}^{-1} \vect{R}_{1}^n$. However, the estimates are generally non-parallel. For example, if $\vect{R}_{1}^n$ is invertible, then
\begin{align}\label{eq:channel_estimates_relation}
\hat{\vect{h}}_{2}^n = \vect{R}_{2}^n{\left(\vect{R}_{1}^n\right)}^{-1}\hat{\vect{h}}_{1}^n.
\end{align}
Notice that the $M\times M$ matrix $\vect{R}_{k}^n
 {({\bf{Q}}_{\rm{tr}}^n)}^{-1}$ for $k=1,2$ only depends on the statistics of the channels $\{{\bf h}_1^n,{\bf h}_2^n\}$. Therefore, it can be precomputed at BS $n$ and only updated when the channel statistics have changed substantially (e.g., due to UE mobility or new scheduling decisions). Therefore, the
MMSE estimator at BS $n$ requires to first compute $\frac{1}{ \sqrt{\rho_{\rm{tr}}} }\vect{Y}^{n} \bphiu^{*}$, which represents the least-square (LS) estimate (e.g., \cite[Sec. 3.4.1]{massivemimobook}) of  $\{{\bf h}_1^n,{\bf h}_2^n\}$, and then multiply it with the precomputed statistical matrix $\vect{R}_{k}^n
 {({\bf{Q}}_{\rm{tr}}^n)}^{-1}$ of each UE. This requires a total number of $NM\tau_p + 2(NM)^2$ complex multiplications per coherence block for all BSs.



\section{Asymptotic SE With Global CSI At All BSs}
We begin by considering a fully cooperative network in which the channel estimates $\{\hat{\bf h}_k^n:\forall n,k\}$ are exchanged among BSs. Since MMSE channel estimation is used, the tightest available lower bound on the ergodic capacity of UE~$1$ is \cite[Th.~4.1]{massivemimobook}
\begin{align} \label{eq:SE-uplink-twousers}
\mathsf{SE}_{1} = \left( 1 - \frac{\taupu}{\tau_c}  \right) {\mathbb{E}}\left\{\log_2  \left( 1 + \gamma_{1}  \right)\right\} \quad \textrm{[bit/s/Hz] }
\end{align}
where $\tau_c$ is number of samples per channel coherence block and $\gamma_{1}$ is the effective SINR given in \eqref{eq:gamma1_UL}, on the top of this page, with
\setcounter{equation}{7} 
\begin{align}\label{Z^n}
{\bf Z}^n = \sum_{k=1}^{2} \left(\vect{R}_{k}^n - \vect{R}_{k}^n {({\bf{Q}}_{\rm{tr}}^n)}^{-1}\vect{R}_{k}^n\right) + \frac{1}{\rho}  \vect{I}_M.
\end{align} 
We define 
\begin{align}
{\bf v}_1 &\triangleq \left[{({\bf v}_1^1)}^{\Ttran},{({\bf v}_1^2)}^{\Ttran},\ldots,{({\bf v}_1^N)}^{\Ttran}\right]^{\Ttran}\in \mathbb{C}^{NM}\\ \hat{\bf h}_k &\triangleq \left[{({\bf h}_k^1)}^{\Ttran},{({\bf h}_k^2)}^{\Ttran},\ldots,{({\bf h}_k^N)}^{\Ttran}\right]^{\Ttran}\in \mathbb{C}^{NM}
\end{align}
for $k=1,2$. The optimal ${\bf v}_1$ that maximizes the SINR expression provided in \eqref{eq:gamma1_UL} is \emph{MMSE combining}~\cite{BHS18A}, given by
\begin{align}\label{eq:SINR_maximizer}
{\bf v}_1=\check{\bf v}_1 & \triangleq \bigg(\hat{\bf h}_2\hat{\bf h}_2^{\Htran} + {\bf Z}\bigg)^{-1}\hat{\bf h}_1
\end{align}
with ${\bf Z} = {{\rm{diag}}\left({\bf Z}^1,{\bf Z}^2,\ldots,{\bf Z}^N\right)} \in \mathbb{C}^{NM\times NM}$ being block-diagonal.
Plugging \eqref{eq:SINR_maximizer} into \eqref{eq:gamma1_UL} yields
\begin{align} \label{eq:gamma1_MMSE}
\gamma_{1}  &=  \hat{\vect{h}}_{1}^{\Htran}\left(  \hat{\vect{h}}_{2} \hat{\vect{h}}_{2}^{\Htran} + {\bf Z}\right)^{-1} \hat{\vect{h}}_{1}.\end{align} 
Notice that MMSE combining requires first the computation of the $MN \times MN$ matrix inverse in \eqref{eq:SINR_maximizer} and then a matrix-vector multiplication. The complexity is also affected by the need of computing the MMSE channel estimates in \eqref{eq:MMSEestimator_h}. Table~\ref{tab:cost_linear_processing} summarizes the total complexity of \eqref{eq:SINR_maximizer} (in terms of number of complex multiplications per coherence block), as obtained from \cite[Sec. 4.1.2]{massivemimobook} under the assumption that the statistical matrices $\bf Z$ and $\{\vect{R}_{1}^n
 {({\bf{Q}}_{\rm{tr}}^n)}^{-1},\vect{R}_{2}^n
 {({\bf{Q}}_{\rm{tr}}^n)}^{-1}: n=1,\ldots,N\}$ are precomputed and stored at BSs.  

\subsection{Asymptotic Analysis for $M \to \infty$ with Fixed $N$}
We will now analyze the asymptotic behavior of $\gamma_{1}$ in \eqref{eq:gamma1_MMSE}
as $M \to \infty$ with $N$ fixed. We assume that the \emph{global} covariance matrices $\{\vect{R}_{k}: k=1,2\}$ are asymptotically linearly independent, which is analytically defined as follows.
\begin{assumption}\label{assumption_4} For $\boldsymbol{\lambda} = [{\lambda}_1, {\lambda}_2]^{\Ttran} \in \mathbb{R}^{2}$ and $i=1,2$, 
\begin{align}\label{eq:condition_CoMP}
\mathop {\liminf}\limits_M \inf_{\{\boldsymbol{\lambda}: \, {\lambda}_i=1\}} \frac{1}{{M}} \left\| \lambda_1 \vect{R}_{1} + \lambda_2 \vect{R}_{2} \right\|_F^2 > 0.
\end{align}
\end{assumption}
This assumption is physically motivated in \cite{BHS18A} and was implicitly made already in \cite{ashikhmin2012pilot}. Under this condition, the following lemma follows.

\begin{table}[t]  
        \caption{Number of complex multiplications per coherence block of different receive combining schemes\vspace{-.2cm}}  \label{tab:cost_linear_processing}
\centering
    \begin{tabular}{|c|c|c|} 
    \hline
    {Scheme} & {Channel estimation}&{Combiner computation}   \\      \hline\hline 
    
    MMSE  &   $NM\tau_p + 2(MN)^2 $ & $3(MN)^2 + MN + \frac{(MN)^3-MN}{3}$   \\   \hline
    
%
    
    D-MMSE  &   $NM\tau_p $ & $ 2NM^2$  \\ \hline

    OBE  &   $NM\tau_p $ & $ 2NM^2$
  
   \\ \hline
  
    \end{tabular}
    \vspace{-0.5cm}
\end{table}

\begin{lemma}~\cite[Th. 1]{BHS18A} \label{theorem:MMSE}
If MMSE combining is used, then under Assumption \ref{assumption_4}, $\gamma_{1}$ increases a.s.~unboundedly as $M\to \infty$. Hence, $\mathsf{SE}_{1}$ increases unboundedly as $M\to \infty$.
\end{lemma}
\begin{IEEEproof}
The proof follows easily from~\cite[App. B]{BHS18A}, but is included since later proofs rely on the expressions defined below. By exploiting the block-diagonal structure of ${\bf Z}$ in \cite[Eq. (43)]{BHS18A}, we obtain
\begin{align}\label{gamma_1_asymptotic_1}
\frac{\gamma_{1}}{M}  \asymp \sum_{n=1}^N \beta_{11}^n - \frac{\big|\sum_{n=1}^N \beta_{12}^n\big|^2}{\sum_{n=1}^N \beta_{22}^n}
\end{align}
where the quantities $\beta_{jk}^n$ are defined as
\begin{align}\label{beta_{jk}^i}
\beta_{jk}^n \triangleq \frac{1}{M}\tr \big({\bf R}_j^n{{({\bf Q}_{\rm{tr}}^n)}^{-1}}{\bf R}_k^n{{({\bf Z}^n)}^{-1}}\big).\end{align}
It can then be proved that
\begin{align}
\liminf_M \, \sum_{n=1}^N\beta_{11}^n  - \frac{{\big|\sum_{n=1}^N\beta_{12}^n\big|} ^2}{ \sum_{n=1}^N\beta_{22}^n } > 0
\end{align}
if Assumption~\ref{assumption_4} is satisfied~\cite{BHS18A}.
\end{IEEEproof}
A similar result can be obtained for UE 2. Hence, the UEs achieve unbounded asymptotic SEs simultaneously, despite pilot contamination. Since the SE is a lower bound on the capacity, we conclude that the asymptotic capacity is unbounded in fully cooperative distributed Massive MIMO networks.

\begin{remark}Assume now that the following condition is satisfied.

\begin{assumption}\label{assumption_1} For $\boldsymbol{\lambda} = [\lambda_1, \lambda_2]^{\Ttran} \in \mathbb{R}^2$ and $i=1,2$, there exists at least one BS $j\in\{1,\ldots,N\}$ for which
\begin{align} \label{eq:assumption_1}
\mathop {\liminf}\limits_M \inf_{\{\boldsymbol{\lambda}: \, \lambda_i=1\}}  \frac{1}{{M}}  \left\| \lambda_1 \vect{R}_{1}^j + \lambda_2 \vect{R}_{2}^j \right\|_F^2 > 0.
\end{align}
\end{assumption} 
Assumption~\ref{assumption_1} is more restrictive than Assumption~\ref{assumption_4}. More precisely, Assumption~\ref{assumption_4} generally holds simply because the BSs have different pathlosses to the UEs, while Assumption~\ref{assumption_1} requires the UEs to have asymmetric spatial channel correlation to one of the BSs. Since the pilot contamination precoding scheme in \cite{ashikhmin2012pilot} is designed for uncorrelated channels, it implicitly relies on Assumption~\ref{assumption_4}. On the other hand, by using the results in~\cite{BHS18A} it easily follows that if Assumption~\ref{assumption_1} holds, then MMSE combining achieves unbounded capacity.
\end{remark}

\begin{remark}[On the physical limits and unbounded capacity] Notice that physics prevent us from letting the size of the array grow indefinitely as $M\to\infty$ and from collecting more energy than it was transmitted.\footnote{Channel gains in cellular communications typically range from $-60$ dB to $-120$ dB. This implies that more than one million antennas are needed to collect more energy than was transmitted. Therefore, this is not an issue when hundreds or thousands of antennas are considered.} Although the limit $M\to \infty$ is not physically achievable, the asymptotic analysis is still an analytical tool to understand what happens at practically large antenna numbers. Indeed, the numerical results of Section~\ref{sectionVI} show that, for practical networks with finite numbers of antennas, the SE with MMSE increases with $M$ and largely outperforms that achieved with the classical maximum-ratio (MR) combining scheme.
\end{remark}


\section{Asymptotic SE in a Distributed Network: Local CSI and Global Statistical Knowledge}\label{sectionIII}

Although MMSE combining is optimal and achieves an unbounded capacity, it requires a fully cooperative network in which the BSs share their channel estimates. In practice, this is hard to achieve since the channels change rapidly over time. Moreover, it has high computational complexity. We now show that a similar scaling behavior can be achieved in a distributed manner with much lower complexity by simply exchanging functionals of the global channel statistics $\{{\bf R}_k: k=1,2\}$.

By utilizing the matrix inversion lemma and the block-diagonal structure of ${\bf Z}$, we can express the MMSE combining  in \eqref{eq:SINR_maximizer} as $\check{\bf v}_1=[(\check{\bf v}_1^1)^{\Ttran} \, \ldots (\check{\bf v}_1^N)^{\Ttran}]^{\Ttran}$ with BS $n$ using
\begin{align}\label{eq:optimal_MMSE}
\!\!\!\check{\bf v}_1^n \!= \!{{({\bf Z}^n)}^{-1}} \left(\hat{\bf h}_1^n - \frac{\sum_{i=1}^N\frac{1}{M} {(\hat{\bf h}_2^i)}^{\Htran}{{({\bf Z}^i)}^{-1}}\hat{\bf h}_1^i}{\frac{1}{M} + \sum_{i=1}^N\frac{1}{M} {(\hat{\bf h}_2^i)}^{\Htran}{{({\bf Z}^i)}^{-1}}\hat{\bf h}_2^i} \hat{\bf h}_2^n\right)\!
\end{align}
where we have multiplied and divided the scaling factor in front of $\hat{\bf h}_2^n$ by $M$. We now exploit the fact that~\cite[App. B]{BHS18A}
\begin{align}
\frac{1}{M} {(\hat{\bf h}_2^n)}^{\Htran}{{({\bf Z}^n)}^{-1}}\hat{\bf h}_1^n &\asymp\beta_{12}^n\\
\frac{1}{M} {(\hat{\bf h}_2^n)}^{\Htran}{{({\bf Z}^n)}^{-1}}\hat{\bf h}_1^n &\asymp\beta_{22}^n
\end{align}
with $\beta_{jk}^n$ given by \eqref{beta_{jk}^i} to propose the alternative  scheme 
\begin{align}\label{eq:MMSE_heuristic}
\vect{\bar v}_1^n \triangleq {{({\bf Z}^n)}^{-1}} \left(\hat{\bf h}_1^n - \frac{\sum_{i=1}^N\beta_{12}^i}{\frac{1}{M} + \sum_{i=1}^N\beta_{22}^i} \hat{\bf h}_2^n\right).\end{align}
This combining vector is a linear combination of the \emph{local} MMSE channel estimates $\{\hat{\bf h}_1^n,\hat{\bf h}_2^n\}$ (with scalar coefficients that depend only on the global channel statistics), followed by a linear transformation with ${{({\bf Z}^n)}^{-1}}$. Therefore, $\vect{\bar v}_1^n$ can be implemented in a distributed manner.
Since in the limiting regime $M\to \infty$, it holds that
\begin{align}\label{eq:asymptotic_MMSE_combining}
\vect{\bar v}_1^n \asymp  \check{\vect{v}}_1^n \asymp {{({\bf Z}^n)}^{-1}} \!\!\left(\hat{\bf h}_1^n - \frac{\sum_{i=1}^N\beta_{12}^i}{\sum_{i=1}^N\beta_{22}^i} \hat{\bf h}_2^n\right)\end{align}
we conclude that $\vect{\bar v}_1^n$ in \eqref{eq:MMSE_heuristic} also achieves unbounded capacity when $M\to \infty$. We call $\vect{\bar v}_1^n$ the \emph{distributed MMSE} (D-MMSE) combining scheme, since it only uses local CSI and global channel statistics.
To get further insights into its structure and computational complexity, we use  \eqref{eq:MMSEestimator_h}--\eqref{eq:channel_estimates_relation} to obtain
\begin{align}\label{eq:v_1_LS}
\vect{\bar v}_1^n= {\bf \Sigma}_1^n\underbrace{\left(\frac{1}{ \sqrt{\rho_{\rm{tr}}} }\vect{Y}^{n} \bphiu^{*}\right)}_{\text{LS channel estimate}}
\end{align}
where
\begin{align}
{\bf \Sigma}_1^n \triangleq {{({\bf Z}^n)}^{-1}} \!\!\left({\bf R}_1^n - \frac{\sum_{i=1}^N\beta_{12}^i}{\frac{1}{M} + \sum_{i=1}^N\beta_{22}^i} {\bf R}_2^n\right)  {({\bf{Q}}_{\rm{tr}}^n)}^{-1}.
\end{align}
To compute $\vect{\bar v}_1^n$ in \eqref{eq:v_1_LS}, BS $n$ must compute the LS channel estimate (by correlating the received 
pilot signal $\vect{Y}^{n} $ with the pilot sequence $\bphiu$)
and then multiply it with the matrix ${\bf \Sigma}_1^n$. Under the assumption that the matrices $\{{\bf \Sigma}_1^n,{\bf \Sigma}_2^n\}$ are available at BS $n$, the total computational complexity of D-MMSE for both UEs is summarized in Table~\ref{tab:cost_linear_processing}. Importantly, it scales as $NM^2$, rather than as $(NM)^3$ as with MMSE combining. Since it is derived to be asymptotically equivalent to MMSE combining (as proved in \eqref{eq:asymptotic_MMSE_combining}), D-MMSE performs better than other distributed schemes with similar complexity; such as the one derived next by following the procedure of~\cite{Neumann2018}.


\section{Asymptotic SE in a Distributed Network \\ Using the Use-and-Then-Forget-Bound}

Inspired by the recent work in~\cite{Neumann2018} for classical Massive MIMO systems (i.e., $N=1$), we now assume that $\vect{v}_k^n$ is obtained as a transformation of the LS channel estimate:
\begin{align} \label{v_k_MR}
\vect{v}_k^n= {\bf W}_k^n\underbrace{\left(\frac{1}{ \sqrt{\rho_{\rm{tr}}} }\vect{Y}^{n} \bphiu^{*}\right)}_{\text{LS channel estimate}}
\end{align}
where ${\bf W}_k^n$ is an arbitrary deterministic matrix that can be optimized.
 Notice that if ${\bf W}_k^n = \vect{R}_{k}^n
 {({\bf{Q}}_{\rm{tr}}^n)}^{-1}$ then $\vect{v}_k^n$ reduces to classical MR combining, based on the MMSE channel estimates, whose SE is known to be asymptotically limited by pilot contamination~\cite{massivemimobook}. Interestingly, \cite{Neumann2018} shows that if ${\bf W}_k^n$ is \emph{optimally} designed at BS $n$ to maximize another  capacity bound, known as the use-and-then-forget (UatF) bound (e.g., \cite[Th.~4.4]{massivemimobook}), then $\vect{v}_k^n$ in \eqref{v_k_MR} achieves unlimited capacity as $M\to \infty$. This is the same scaling behaviour as for MMSE combining, but there will anyway be a performance gap. Next, we first revisit the approach in \cite{Neumann2018} for the investigated distributed Massive MIMO setup and then provide insights into the reason behind its scaling behaviour.
 
 \begin{figure*}[ht] \vspace{-0.1cm}
 \setcounter{equation}{34}
\begin{align}\label{eq:gamma_1.1}
\frac{\gamma_1^{\rm{UatF}}}{M} = \sum\limits_{n=1}^N\frac{1}{M}{({\bf r}_1^n)}^{\Htran}{({\bf Q}^n)}^{-1}{\vect{r}}_{1}^n - \frac{\sum\limits_{n=1}^N\Big|\frac{1}{M}{({\bf r}_1^n)}^{\Htran}{({\bf Q}^n)}^{-1}{\vect{r}}_{2}^n\Big|^2}{\frac{1}{M}+\frac{1}{M}\sum\limits_{n=1}^N{({\bf r}_2^n)}^{\Htran}{({\bf Q}^n)}^{-1}{\vect{r}}_{2}^n}  = \sum\limits_{n=1}^N\alpha_{11}^n  - \frac{\left|{\sum_{n=1}^N\alpha_{12}^n} \right|^2}{ \frac{1}{M} + \sum_{n=1}^N\alpha_{22}^n }\triangleq \delta_1
\end{align}\vspace{-0.1cm}
\hrule\vspace{-0.5cm}
\end{figure*}

  \setcounter{equation}{25}
 \subsection{Revisiting \cite{Neumann2018} for Distributed Massive MIMO}
 The bound provided in \eqref{eq:SE-uplink-twousers} requires the use of MMSE channel estimation \cite[Th. 4.1]{massivemimobook}. Therefore, it cannot be applied with \eqref{v_k_MR}. On the contrary, the UatF bound can be applied along with any channel estimator \cite[Th. 4.4]{massivemimobook}. By using it, the capacity of UE~$1$ can be lower bounded by
\begin{align} \label{eq:SE-uplink-twousers-UatF}
\mathsf{SE}_{1}^{\rm UatF} = \left( 1 - \frac{\taupu}{\tau_c}  \right) \log_2  \left( 1 + \gamma_{1}^{\rm UatF}  \right) \quad \textrm{[bit/s/Hz] }
\end{align}
where the effective SINR $\gamma_{1}^{\rm UatF}$ is (e.g.~\cite[Eq.~(24)]{BSD16A})  
\begin{align}\label{eq:approximationOfGamma}
\!\!\!\!\gamma_{1}^{\rm{UatF}}
\!\!\!= \!\frac{\left|\sum\limits_{n=1}^N\tr ( {(\vect{W}_1^n)}^{\Htran}\vect{R}_{1}^n)\right|^2}{\!\!\!\!\left|\sum\limits_{n=1}^N\tr \big( {(\vect{W}_1^n)}^{\Htran}\vect{R}_{2}^n \big)\right|^2 \!\!+\!  \sum\limits_{n=1}^N\tr \big({(\vect{W}_1^n)}^{\Htran}\vect{Q}_{\rm{tr}}^n \vect{W}_1^n\vect{Q}^n\big)}\!\!\!
\end{align}
with 
\begin{align}
{\bf{Q}}^n = \vect{R}_{1}^n +  \vect{R}_{2}^n + \frac{1}{ \rho} \vect{I}_{M}.
\end{align}
Notice that ${\bf{Q}}^n$ is equal to ${\bf{Q}}_{\rm{tr}}^n$ in \eqref{eq:covariance_matrix_pilot_phase} only if the same SNR is imposed for pilot and data transmissions, i.e., $\tau_p=1$. In practice, however, it is common to have $\tau_p>1$ to achieve good estimation quality also for UEs with weak SNRs. 

We now look for the matrices $\{\vect{W}_1^n: n=1,\ldots,N\}$ that maximize $\mathsf{SE}_{1}^{\rm UatF}$ with $\gamma_{1}^{\rm{UatF}}$ given by \eqref{eq:approximationOfGamma}. To this end, we define ${\bf w}_1^n = {\vectorize({\bf W}_1^n}) \in \mathbb{C}^{M^2}$ and ${\vect{r}}_{i}^n = {\vectorize({\bf R}_{i}^n}) \in \mathbb{C}^{M^2}$
for $i=1,2$. Then, similarly to \eqref{eq:gamma1_UL} we rewrite \eqref{eq:approximationOfGamma} as follows\footnote{We use
$\tr ({(\vect{W}_1^n)}^{\Htran}\vect{Q}_{\rm{tr}}^n \vect{W}_1^n\vect{Q}^n) 
= {({\bf w}_1^n)}^{\Htran} ({(\vect{Q}_{\rm{tr}}^n)}^{\Ttran} \otimes {\bf Q}^n){\bf w}_1^n$.}
\begin{align}\label{eq:gamma_1}
\!\gamma_{1}^{\rm{UatF}}\!
&= \!\frac{\left|\sum\limits_{n=1}^N{({\bf w}_1^n)}^{\Htran}{\bf r}_{1}^n\right|^2}{\left|\sum\limits_{n=1}^N{({\bf w}_1^n)}^{\Htran}{\bf r}_{2}^n\right|^2 \!+ \! \sum\limits_{n=1}^N{(\vect{w}_1^n)}^{\Htran}{\bf U}^n \vect{w}_1^n\!} 
\end{align}
with $
{\bf U}^n = {(\vect{Q}_{\rm{tr}}^n)}^{\Ttran} \otimes {\bf Q}^n\in \mathbb{C}^{M^2\times M^2}$. Let us further define
\begin{align}
{\bf w}_1 &= {\left[{({\bf w}_1^1)}^{\Ttran},{({\bf w}_1^2)}^{\Ttran},\ldots,{({\bf w}_1^N)}^{\Ttran}\right]}^{\Ttran}\in \mathbb{C}^{NM^2}\\{\bf r}_{k} &= {\left[{({\bf r}_k^1)}^{\Ttran},,{({\bf r}_k^2)}^{\Ttran}\ldots,{({\bf r}_k^N)}^{\Ttran}\right]}^{\Ttran} \in \mathbb{C}^{NM^2}
\end{align}
for $k=1,2$ and ${\bf U} = {{\rm{diag}}\left({\bf U}^1,{\bf U}^2,\ldots,{\bf U}^N\right)} \in \mathbb{C}^{NM^2\times NM^2}$. Then, $\gamma_{1}^{\rm{UatF}}$ in \eqref{eq:gamma_1} can be rewritten as 
\begin{align}\label{eq:gamma_1.UatF}
\gamma_{1}^{\rm{UatF}} = \frac{\left|{\bf w}_1^{\Htran}{\bf r}_{1}\right|^2}{\left|{\bf w}_1^{\Htran}{\bf r}_{2}\right|^2 + {\bf w}_1^{\Htran}{\bf U}{\bf w}_1}=\frac{\left|{\bf w}_1^{\Htran}{\bf r}_{1}\right|^2}{{\bf w}_1^{\Htran} \left({\bf r}_{2}{\bf r}_{2}^{\Htran} + {\bf U}\right){\bf w}_1}
\end{align}
which is a generalized Rayleigh quotient with respect to ${\bf w}_1$. The maximum is thus achieved by  
\begin{align}\label{optimal_a}
{\bf w}_1 = \left({\bf r}_{2}{\bf r}_{2}^{\Htran} + {\bf U}\right)^{-1}{\bf r}_{1}.
\end{align}
Plugging the above result into \eqref{eq:gamma_1.UatF} yields
\begin{align}
\!\!\!\!\gamma_{1}^{\rm{UatF}} 
 = {\bf r}_{1}^{\Htran}\left({\bf r}_{2}{\bf r}_{2}^{\Htran} + {\bf U}\right)^{-1}\!{\bf r}_{1}\label{eq:gamma_1_opt}.
\end{align}
By substituting ${\bf W}_1 = \vectorize^{-1}({\bf w}_1)$ into \eqref{v_k_MR} leads to what~\cite{Neumann2018} calls \emph{optimal bilinear equalizer (OBE)}. The computational complexity of OBE is reported in Table~\ref{tab:cost_linear_processing} and coincides with that of D-MMSE, as it easily follows from \eqref{eq:v_1_LS} and \eqref{v_k_MR}.

\subsection{Asymptotic Analysis of $\gamma_1^{{\rm{UatF}}}$ for $M \to \infty$ with Fixed $N$}
As done for $\gamma_{1}$, we now analyze the asymptotic behavior of $\gamma_{1}^{{\rm{UatF}}}$ in \eqref{eq:gamma_1_opt} when $M \to \infty$.

\begin{lemma} \cite{Neumann2018} \label{theorem:GMF}
If OBE is used, then under Assumption \ref{assumption_4}, $\gamma_1^{\rm{UatF}}$ and $\mathsf{SE}_{1}^{\rm{UatF}}$ increase unboundedly as $M\to \infty$.
\end{lemma}
\begin{IEEEproof}
 \setcounter{equation}{35}
By applying the matrix inversion lemma and by also multiplying and dividing each term by $M$, we may rewrite $\gamma_1^{\rm{UatF}}$ in \eqref{eq:gamma_1_opt} as \eqref{eq:gamma_1.1} on the top of the page where the quantities $\alpha_{jk}^n$ are defined as, for $n=1,\ldots,N$,
\begin{align}\label{alpha_{jk}^i}
\alpha_{jk}^n \triangleq  \frac{1}{M}\tr \big({\bf R}_j^i{{({\bf Q}_{\rm{tr}}^n)}^{-1}}{\bf R}_k^i{{({\bf Q}^n)}^{-1}}\big).\end{align}
Notice that \eqref{eq:gamma_1.1} holds for any $M$, not only for $M\to\infty$.
From \cite[Th. 1]{BHS18A}, we have that, under Assumption~\ref{assumption_4}, $\liminf_M\delta_1 >0$. Therefore, $\gamma_1^{\rm{UatF}} $ grows unboundedly with $M$. 
\end{IEEEproof}
A similar result can be proved for UE~2 by interchanging indices, thus OBE achieves unbounded capacity as $M\to \infty$. 

\subsection{Interpretation and Key Insights}
We now provide an explicit form of OBE that provides an intuitive interpretation and explanation of the result, which were missing in~\cite{Neumann2018}. As shown in the appendix, ${\bf W}_1 = \vectorize^{-1}({\bf w}_1)$ is explicitly given by
\begin{align}\label{eq:sectiond.5}
\!\!{\bf W}_1^n & = {({\bf Q}^n)}^{-1}\left({\bf R}_{1}^n - \frac{\sum_{i=1}^N\alpha_{12}^i}{\frac{1}{M} + \sum_{i=1}^N\alpha_{22}^i}{\bf R}_{2}^n\right){({\bf Q}_{\rm{tr}}^n)}^{-1}.
\end{align}
Plugging \eqref{eq:sectiond.5} into \eqref{v_k_MR} yields
\begin{align}\label{eq:sectiond.5.1}
\vect{v}_1^n & = {({\bf Q}^n)}^{-1} \left(\hat{\bf {h}}_{1}^n - \frac{\sum_{i=1}^N\alpha_{12}^i}{\frac{1}{M} + \sum_{i=1}^N\alpha_{22}^i} \hat{\bf {h}}_{2}^n\right)
\end{align}
where we have used \eqref{eq:MMSEestimator_h} and \eqref{eq:channel_estimates_relation}.
Similarly to D-MMSE in \eqref{eq:MMSE_heuristic}, the explicit form in \eqref{eq:sectiond.5.1} reveals that the optimal (in the sense of maximizing $\gamma_{1}^{\rm{UatF}}$ in \eqref{eq:gamma_1}) OBE of BS $n$ is obtained as a linear combination of its local MMSE channel estimates $\{\hat{\bf {h}}_{1}^n,\hat{\bf {h}}_{2}^n\}$, followed by a linear transformation with ${{({\bf Q}^n)}^{-1}}$.
The key differences with respect to D-MMSE in \eqref{eq:MMSE_heuristic} are the matrices used in the linear transformation and in the computation of the scalar coefficients $\{\alpha_{jk}^i\}$ in front of $\hat{\bf {h}}_{2}^n$. Since our new formulation shows that OBE eventually requires MMSE channel estimation, it can also be used with the tighter bound in \eqref{eq:SE-uplink-twousers}. In the next section, we show that it provides much better performance than predicted by the UatF bound, which vastly underestimates the SE \cite[Sec. 4.2.1]{massivemimobook}. 

\begin{remark}Interestingly, the explicit form in \eqref{eq:sectiond.5.1} resembles that obtained in \cite{JorswieckLD08} for the achievable rate region in the downlink of multiple-input single-output (MISO) interference channels, under the assumption of perfect CSI. In particular, the authors showed that, if ${\bf v}^n \in \mathbb{C}^M$ is a precoding vector that gives a point on the Pareto boundary of the rate region, then it can be expressed as a linear combination of the linearly transformed true channel vectors $\{{\bf h}_{1}^n,{\bf h}_{2}^n\}$. 
\end{remark}

\begin{figure}[t!]
 \centering\vspace{-0.1cm}
\includegraphics[width=.43\columnwidth]{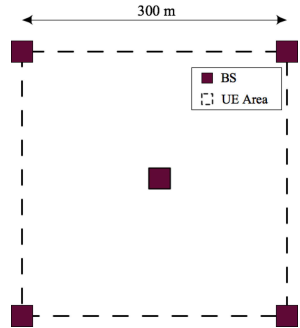}
\vspace{-0.3cm}
\caption{A distributed Massive MIMO setup with UEs uniformly and independently distributed in the area.} \label{figureSetup}  \vspace{-0.2cm}
\end{figure}

\section{Numerical Results}\label{sectionVI}

We consider the setup in Fig.~\ref{figureSetup} with the $K=2$ UEs uniformly and independently distributed in the area, with a minimum distance from BSs of $20$ m. Results are obtained by averaging over $100$ UE locations. We consider the exponential correlation model for a uniform linear array with large-scale fading $\varsigma_k^n=\tr ( \vect{R}_{k}^{n} )/M$. This leads to 
$
[ \vect{R}_k^n]_{m,n} = \varsigma_k^n r^{|n-m|} e^{\imath  (n-m) \theta_k^n}
$
where $r = 0.5$ is the correlation factor and  $\theta_k^n$ is the angle-of-arrival from UE $k$ to BS $n$ \cite{BHS18A}. 
We consider pilots with $\tau_p = 10$ and coherence blocks of $\tau_c = 200$ channel uses. The average SNR observed at a BS antenna for data transmission is $\rho \tr({\vect{R}_{k}^{n}})/N = 6.6$ dB.
\begin{figure}[t!]
\centering
\includegraphics[width=1.12\columnwidth]{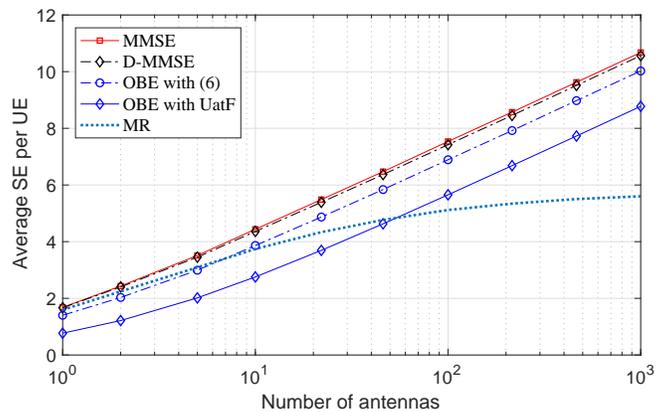}
\vspace{-0.5cm}
\caption{Average SE per UE [bit/s/Hz/UE] as a function of $M$, for covariance matrices based on the exponential correlation model and $K=2$.\vspace{-0.5cm}} \label{figureAntennas} 
\end{figure}

The average SE per UE is shown in Fig.~\ref{figureAntennas} with the MMSE, D-MMSE, and OBE schemes. The latter is used with the UatF bound and with \eqref{eq:SE-uplink-twousers} since it was shown to be obtained as a linear combination of local MMSE channel estimates. These schemes are compared with the classical MR combining, based on MMSE channel estimates. Fig.~\ref{figureAntennas} shows that the SEs of the three investigated schemes grow without bound as $M \to \infty$ while MR converges to a finite limit. MMSE provides the highest SE, since it is optimal. The loss incurred by using D-MMSE is negligible whereas it is around 6--16\% for OBE using the capacity bound \eqref{eq:SE-uplink-twousers}. This is quite remarkable for combiners that make only use of local CSI. While D-MMSE approaches MMSE as $M\to \infty$ (since it is derived to be asymptotically equivalent to MMSE combining), there is no corresponding result for OBE with \eqref{eq:SE-uplink-twousers} since OBE was optimally derived and proved to achieve unbounded asymptotic capacity only with the UatF bound in \eqref{eq:SE-uplink-twousers-UatF}. When using this bound, OBE performs worse (even worse than MR for $M\le 50$), showing that \eqref{eq:SE-uplink-twousers-UatF} vastly underestimates the SE.

\section{Conclusions}
We considered distributed Massive MIMO and investigated how to achieve an asymptotic unbounded SE in the uplink of a two-user network, with spatial correlated channels, pilot-contamination, and different degrees of CSI. Based on the optimal MMSE combining with full CSI, a new asymptotically optimal scheme was derived to achieve unbounded SE using only local CSI at each BS and global channel statistics. We also provided key insights into the generalization of a known OBE scheme, which was shown to achieve good performance, despite being derived on the basis of an overly conservative capacity lower bound. Due to space limitations, we were not able to consider a distributed Massive MIMO network with $K>2$ UEs. However, we anticipate that similar conclusions hold for this case as well. Particularly, the D-MMSE combining vector $\vect{\bar v}_k^n$ of UE $k$ can be computed as follows:
\begin{align}\label{eq:optimal_DMMSE_1}
\vect{\bar v}_k^n = {({\bf Z}^n)}^{-1} \left(\sum_{i=1}^K\varsigma_{ki}\hat{\bf h}_i^n\right)\!
\end{align}
where 
${\boldsymbol{\varsigma}}_k = \big({\bf B}+ {1}/{M}{\bf I}_K\big)^{-1}{\bf e}_k$
with ${\bf B}\in \mathbb{C}^{K\times K}$ and $\big[{\bf B}\big]_{\ell j}\triangleq \sum_{n=1}^N \beta_{j \ell} ^n$, with $\beta_{j \ell} ^n$ given by \eqref{beta_{jk}^i}. As in the two-user case, $\vect{\bar v}_k^n$ is obtained as a linear combination of the local channel estimates $\{\hat{\bf h}_1^n, \ldots, \hat{\bf h}_K^n\}$ with scalar coefficients that depend only on the global channel statistics and achieves an unbounded capacity when $M \to \infty$.
In the extended version, the technical details for an arbitrarily large number of UEs will be provided and the asymptotic analysis will be extended to the case in which the number of BSs grows infinitely large.

\section*{Appendix}

Define by ${\bf R} \!=\! [{\bf r}_{1},{\bf r}_{2} ] \in \mathbb{C}^{NM^2 \times 2}$ the matrix collecting the vectorized covariance matrices $\{{\bf r}_{1},{\bf r}_{2}\}$ as columns. Note that ${\bf w}_1$ in \eqref{optimal_a} can be rewritten as $
{\bf w}_1 = \left(1 + {\bf r}_1^{\Htran} {\bf U}^{-1}{\bf r}_1\right)\left({\bf R}{\bf R}^{\Htran} + {\bf U} \right)^{-1}{\bf r}_1$.
Since the SINR expression in \eqref{eq:gamma_1_opt} does not change if we scale ${\bf w}_1$ by
any non-zero scalar, we can also use 
\begin{align}\label{eq:Appendix_37}
{\bf w}_1 =\left({\bf R}{\bf R}^{\Htran} + {\bf U} \right)^{-1}{\bf R}{\bf e}_1
\end{align}
with ${\bf e}_1$ being the first vector of the canonical basis.
By applying the matrix inversion lemma, we rewrite ${\bf w}_1$ in \eqref{eq:Appendix_37} as $
{\bf w}_1 
= {\bf U}^{-1}{\bf R}{\bf a}$ where 
\begin{align}
{\bf a} = \big({\bf R}^{\Htran}{\bf U}^{-1}{\bf R}+{\bf I}_2\big)^{-1}{\bf e}_1.
\end{align}
By recalling $\vect{U}^n = {(\vect{Q}_{\rm{tr}}^n)}^{\Ttran} \otimes {\bf Q}^n$ and reverting the vectorization (by using $\vectorize({\bf ABC}) = (\vect{C}^{\Ttran} \; \otimes \; {\bf A})\vectorize({\bf B})$), we obtain
\begin{align}
\!\!\!{\bf W}_1 = {\bf Q}^{-1}\left( a_{1}{\bf R}_{1} + a_{2}{\bf R}_{2}\right){\bf Q}_{\rm{tr}}^{-1}\label{eq:sectiond.4}
\end{align}
where we have defined $a_{i} = \left[{\bf a}\right]_{i}$. Notice that 
\begin{align}
\!\!\left[{\bf R}^{\Htran}{\bf U}^{-1}{\bf R}\right]_{j,k}\!\!=\!\!\sum_{n=1}^N{({\bf r}_j^n)}^{\Htran}{({\bf U}^n)}^{-1}{{\bf r}_k^n}  = M\sum_{n=1}^N\alpha_{jk}^n\triangleq M \alpha_{jk}
\end{align}
with $\alpha_{jk}^n$ given by~\eqref{alpha_{jk}^i}.
After simple calculus, we 
obtain
\begin{align}\label{b_1^opt}
\!\!\!\!{\bf a} \!= \!\frac{1}{M}\frac{1}{\big(\frac{1}{M}+\alpha_{11}\big)\big(\frac{1}{M} +\alpha_{22}\big) \!-\! |\alpha_{12}|^2} \left[ \! {\begin{array}{*{20}{c}}
\frac{1}{M} + \alpha_{22}\\
-\alpha_{12}
\end{array}} \! \right]\!\!.
\end{align}
By using \eqref{b_1^opt}, we 
eventually obtain \eqref{eq:sectiond.5} where we have dropped the scaling factor $1/\mu$ with $\mu = {\frac{1}{M}+ \alpha_{11} - \frac{|\alpha_{12}|^2}{\frac{1}{M} + \alpha_{22}}} 
$ since it does affect the SINR in \eqref{eq:gamma_1_opt}.

\bibliographystyle{IEEEtran}
\bibliography{IEEEabrv,ref}

\end{document}